\documentstyle[12pt]{article}

\newcommand{\bbc}{\begin{center}}
\newcommand{\eec}{\end{center}}

\begin{document}
\begin{flushright}
Alberta Thy-14-98 \\
December 1998\\
\end{flushright}

\vspace {0.3in}

\begin{center}
\Large\bf Helicity and partial wave amplitude analysis of $D \rightarrow K^{*} {\rho}$ decay\\
\vspace {0.3in}
{\large El hassan  El aaoud and A. N. Kamal  } \\ \vspace {0.1in}
 \small\em Theoretical Physics Institute and Department of Physics, \\
 \small\em University of Alberta, Edmonton, Alberta Canada T6G 2J1.
\end{center}
\vspace {0.1in}

\begin{abstract}
We have carried out an analysis of helicity and partial-wave amplitudes for the process $D \rightarrow K^{*} {\rho}$ in the factorization approximation using several models for the form factors. All the models, with the exception of one, generate partial-wave amplitudes with the hierarchy $\mid S \mid > \mid P \mid > \mid D \mid$. The one exception gives $\mid S \mid >\mid D \mid >\mid P \mid$. Even though in most models the $D$-wave amplitude is an order of magnitude smaller than the $S$-wave amplitude, its effect on the longitudinal polarization could be as large as $30\%$.  Due to a misidentification of the partial-wave amplitudes in terms of the Lorentz structures in the relevant literature, we cast doubt on the veracity of the listed data, particularly the partial-wave branching ratios. 
\end{abstract}

\bbc (PACS numbers: 13.25.-k, 13.25.Ft) \eec
\newpage

\bbc 
{\bf I. INTRODUCTION}
\eec

The weak hadronic decay of charm $D$ meson to two resonant vector particles is difficult to analyze experimentally, as well as to understand theoretically. At the theoretical level much past effort was devoted to understand mainly the rate $\Gamma(D \rightarrow V_1V_2)$ ( $V$ stands for vector meson). Studies based on factorization model were carried out by Bauer el al.\cite{ref:bauer} and Kamal et al. \cite{ref:kamal}; approaches based on flavor $SU(3)$ symmetry and broken $SU(3)$ symmetry were pursued also by Kamal et al. \cite{ref:kamal} and by Hinchliffe and Kaeding \cite{ref:hin}; and Bedaque et al.\cite{ref:beda} made a pole-dominance model calculation .

One peculiarity of a pseudoscalar meson, $P$, decaying into two vector 
mesons is that the final-state particles are produced in different 
correlated polarization states. The hadronic matrix element, $A = 
\left\langle V_1V_2\mid H_{weak} \mid P\right\rangle$, involves three invariant amplitudes which can be expressed in terms of three different, but equivalent,  bases; the helicity basis $|++\rangle, |--\rangle, |00\rangle$, the partial-wave basis (or the $LS$-basis) $|S\rangle, |P\rangle, |D\rangle$ and the transversity basis $|0\rangle, |\parallel\rangle, |\perp\rangle$. The interrelations between the amplitudes in these bases are presented in the next section. The data \cite{ref:pdg} for $D\rightarrow {K}^{*}\rho$ decay are quoted either in terms of the helicity branching ratios or the partial-wave branching ratios. Hence our study of the process $D \rightarrow {K}^{*}\rho$ is carried out in these two bases. We have undertaken a theoretical analysis for the particular decay, $D \rightarrow K^{*} {\rho}$, assuming factorization (more on it in section 2) and using a variety of models for the form factors. Such a study has not been undertaken in the past.

The experimental analysis of $D \rightarrow K^*{\rho} $ ( measurement of the branching ratio, partial-wave branching ratios, polarization etc.) is done by considering the resonant 
substructure of the four-body decays $D \rightarrow K \pi \pi \pi$ \cite{ref:coff, ref:anj}. There are several two-body decay modes (examples: $D\rightarrow {K}^{*}\rho$ and  $D\rightarrow 
K{a}_{1}$) which contribute to the final states in $D^0 \rightarrow \bar{K}^0 \pi^- \pi^+ \pi^0, D^+ \rightarrow K^- \pi^+ \pi^+ \pi^0, D^0 \rightarrow K^- \pi^+ \pi^+ \pi^- $. Following the standard practice, Refs. \cite{ref:coff, ref:anj}  took the signal terms of the probability density to be a coherent sum of complex amplitudes for each decay chain leading to the four-body decays 
of $D$. Hence, the different contributing amplitudes can interfere among themselves. In general, the interference terms do not integrate to zero (see \cite{ref:adler} for more details about the three-body decay $D \rightarrow K \pi \pi $).  Consequently, the sum of the fractions $f_i$ does not add up to unity: $\sum{f_i} \neq 1$( see ref.\cite{ref:coff, ref:adler, ref:jon}). The branching fractions into two-body channels are then determined by maximizing the likelihood function. The branching fraction into any particular two-body channel, such as $D\rightarrow {K}^{*}\rho$, can be analyzed in terms of the helicity amplitdes (${H}_{++}$, ${H}_{--}$, ${H}_{00}$), or the partial-wave amplitudes ($S$, $P$, $D$), or the transversity amplitudes (${A}_{0}$, ${A}_{\parallel}$, ${A}_{\perp}$). As a results of the completeness of each one of these bases, the decay rate $\Gamma(D \rightarrow {K}^{*}\rho)$ is expressed as an incoherent sum 
of the helicity, or the partial-wave, or the transversity amplitudes \cite{ref:ali, ref:digh, 
ref:gkh}. This imposes some constraints on the helicity and the partial-wave branching fractions $B$ as they should add up to the total branching fraction for the mode $D \rightarrow{K}^{*}\rho$ as follow: $B_{++} + B_{--} + B_{00} = B_0 + B_\parallel +B_\perp = B_S + B_P + B_D = B_{{K}^{*}\rho}$. A similar situation occurs in $\bar{n}p$ annihilation to 3 ( and 5) pions \cite{ref:amado} where $S$ and $P$ waves are treated incoherently. An obvious problem with the ${D}^{0}\rightarrow {{K}^{*}}^{0}{\rho}^{0}$ data \cite{ref:pdg} is that this constraint is violated: the sum of the branching fractions into $S$ and $D$ states exceeds the total branching fraction. The fact that this sum also exceeds the transverse branching fraction is, by itself, not a problem due to the interference between the S and D waves. However, the problem with the data \cite{ref:pdg} is that the transverse branching fraction saturates the total branching fraction. There is, therefore, an internal inconsistency in the data: all the data listings can not be 
correct. The Particle Data Group listing of $D\rightarrow {K}^{*}\rho$ data has remained unchanged since 1992 .

We believe that the source of the inconsistency in the data \cite{ref:pdg,ref:coff,ref:anj} has to do with the identification of the partial-wave amplitudes, $S$, $P$ and $D$, with the Lorentz structures in the decay amplitude (see Table II and, especially, Eqns. (32) - (34) of \cite{ref:coff}). The decay amplitude $A$ for the process $P\rightarrow {V}_{1}{V}_{2}$ is expressed in terms of three independent Lorentz structures and their coefficients, represented in the notation of \cite{ref:gv, ref:krp} by $a, b, c$, and in the notation of  \cite{ref:bsw1} by the form factors $A_1(q^2), A_2(q^2)$ and $V(q^2)$ . We discuss this point in detail in the next section, but suffice it to say here that in \cite{ref:coff} the $P$-wave amplitude is identified with $c$ of \cite{ref:gv,ref:krp} (or $V$ of \cite{ref:bsw1}), which is correct; however, they identify the $S$-wave amplitude with $a$ of \cite{ref:gv,ref:krp} (or ${A}_{1}$ of \cite{ref:bsw1}) and $D$-wave amplitude with $b$ of \cite{ref:gv,ref:krp} (or ${A}_{2}$ of \cite{ref:bsw1}), which is incorrect. We discuss this point in some detail in sections 3 and 4.

Part of the problem could also be that the transverse amplitudes $H_{++}, H_{--}$ and the longitudinal amplitude $H_{00}$ were fitted independently in \cite{ref:coff}. Their argument for doing so was the large measured polarization of $K^*$ in semileptonic decay of the $D$ meson 
\cite{ref:anj89}. However, later measurements \cite{ref:pdg94} of the polarization of $K^*$ being much smaller vitiate this procedure. 

\vskip 5mm  
\begin{center}
{\bf II.  METHOD}
\end{center}
The decay $D \rightarrow K^*{\rho}  $  is Cabibbo-favored and is induced by the effective weak Hamiltonian which can be reduced to the following color-favored ($CF$) and color-suppressed ($CS$) forms \cite{ref:ksuv}:
\begin{equation}
{H }_{CF}=  \frac{G_{F}}{\sqrt{2}} V_{cs} V^{*}_{ud} [ a_{1}( 
\bar{u} d)( \bar{s} c)  + {c}_{2} {O}_{8}],
\end{equation}
\begin{equation}
{H}_{{CS}}=  \frac{G_{F}}{\sqrt{2}} V_{cs} V^{*}_{ud} [ a_{2}( 
\bar{u} c)( \bar{s} d)  + {c}_{1} \tilde{O}_{8}],
\end{equation}
where $V_{qq'}$ are the CKM matrix elements. The brackets $( \bar{u} d)$ represent $( V - A )$ color-singlet Dirac bilinears. ${O}_{8}$ and ${\tilde{O}}_{8}$ are products of color octet currents:  $ O_8 = \frac{1}{2}\sum_{a=1}^{8}{(\bar{u}{\lambda}^{a}d)(\bar{s}{\lambda}^{a}c})$ and ${\tilde{O}}_{8} = \frac{1}{2}\sum_{a=1}^{8}{(\bar{u}{\lambda}^{a}c)(\bar{s}{\lambda}^{a}d})$.  ${\lambda}^{a}$ are the Gell-mann matrices. $a_1$ and $a_2$ are the Wilson coefficients for which we use the values $a_1 =1.26\pm 0.04$ and $a_2 = -0.51 \pm 0.05$ \cite{ref:ksuv} . In general $a_1$ and $a_2$ are related to the coefficients $c_1$ and $c_2$ \cite{ref:neubert} by
\begin{equation}
a_1=c_1+{c_2 \over N}, ~~~~~~a_2=c_2+{c_1 \over N}
\end{equation}
where N is an effective number of colors, and $c_1=1.26$, $c_2=-0.51$ \cite{ref:neubert}. Using a value of N different from $3$ is a way to parametrize nonfactorization effects. Our parametrization amounts to setting $N\rightarrow \infty$. This particular decay, $D \rightarrow K^{*} {\rho}$, has also been studied by Kamal et al. \cite{ref:ksuv} and by Cheng \cite{ref:hyc} from the view point of explicit ( rather than implicit as here) nonfactorization.

 In the factorization approximation one neglects the contribution from ${O}_{8}$ and ${\tilde{O}}_{8}$, and the  matrix element of the first term is written as a product of  two current matrix elements. Since we are effectively working with $N \neq 3$, one could argue that the nonfactorization arising from ${O}_{8}$ and ${\tilde{O}}_{8}$ is being included. We consider the following three decay: (i) $D^0 \rightarrow K^{-*} {\rho}^+  $, a color-favored decay which gets contribution from external ${\it W}$- exchange, known as a class I process, (ii) $D^0 \rightarrow \bar{K}^{0*} {\rho}^0 $ , a color-suppressed process which get contribution from internal ${\it W}$ - exchange, known as a class II process, and (iii)  $D^+ \rightarrow \bar{K}^{0*} {\rho}^+$, a class III process which gets contribution from external as well as internal ${\it W}$-exchange mechanisms. The  decay amplitudes are given by:
\begin{eqnarray}
A(D^0 \rightarrow {K}^{-*} {\rho}^+) &=& {G_F \over \sqrt{2} } 
V_{cs}V_{ud}^* a_1\left\langle\rho^+ \mid\bar{u} d\mid 0 
\right\rangle\left\langle {K}^{-*}\mid \bar{s} c \mid D^0\right\rangle 
. \\
A(D^0 \rightarrow \bar{K}^{0*} {\rho}^0) &=&{G_F \over \sqrt{2} } 
V_{cs}V_{ud}^* {a_2 \over \sqrt{2} }\left\langle\bar{K}^{0*} 
\mid\bar{s} 
d\mid 0 \right\rangle\left\langle \rho^0\mid \bar{u} c \mid 
D^0\right\rangle.
\label{eq:d00} \\
A(D^+ \rightarrow \bar{K}^{0*} {\rho}^+) &=& {G_F \over \sqrt{2} } 
V_{cs}V_{ud}^*\left\{ a_1\left\langle\rho \mid\bar{u} d\mid 0 
\right\rangle\left\langle \bar{K}^{*0}\mid \bar{s} c \mid 
D^+\right\rangle \right. \nonumber\\
 & & \left.+ a_2 \left\langle\bar{K}^{0*} \mid\bar{s} d\mid 0 
\right\rangle\left\langle \rho^+\mid \bar{u} c \mid D^+\right\rangle 
\right\}  \nonumber\\
& = &  A(D^0 \rightarrow {K}^{-*} {\rho}^+) + \sqrt{2} A(D^0 
\rightarrow \bar{K}^{0*} {\rho}^0) 
\end{eqnarray}
the extra $\sqrt{2}$ in Eq.  (\ref{eq:d00}) comes from the flavor part 
of the wave function of ${\rho}^0 $.
Each of the current matrix elements can be expressed in terms of meson 
decay constants and  invariant form factors. We use the following 
definitions:
\begin{eqnarray}
\left\langle V\mid \bar{u} d \mid 0\right\rangle 
&=&m_Vf_V\varepsilon_{\mu}^* \\
\vspace{7mm}
<V \mid\bar{s} c\mid D> &=& {2 \over m_D + m_V} 
{\epsilon}_{\mu\nu\rho\sigma}\varepsilon_V^{*\nu}P_{D}^{\rho}P_V^{\sigma} V(q^2) +i\{\varepsilon_{V \mu}^*(m_D + m_V)A_1(q^2) -  \nonumber\\
\vspace{7mm}
&& {\varepsilon_V^*.q \over m_D + m_V} (P_V + P_D)_\mu A_2(q^2) 
-{\varepsilon_V^*.q \over q^2} 2 m_V q_\mu A_3(q^2) + \nonumber\\
\vspace{7mm}
&&  {\varepsilon_V^*.q \over q^2} 2 m_V q_\mu A_0(q^2) \},
\end{eqnarray}
 where $ q = P_D -P_V$ is the momentum transfer, $f_V$ is the decay 
constant of the vector meson $V$, $\varepsilon_V$ is its polarization,  
$ A_i(q^2), ( i = 1, 2, 3 )$ and $V(q^2)$ are invariant form factors 
defined in \cite{ref:bsw1} . In terms of the helicity amplitudes the 
decay rate is given by
\begin{equation}
\Gamma(D \rightarrow V_1 V_2 ) = {{\it p} \over 8\pi m_D^2}\left\{ 
|H_{00}|^2 + |H_{++}|^2 + |H_{--}|^2 \right\}.
\label{eq:dec}
\end{equation}
where {\it p} is the center of mass momentum in the final state. 
$H_{00}, H_{++} $ and $H_{--}$ are the longitudinal and the two 
transverse helicity amplitudes, respectively, for the decay $D^0 \rightarrow 
{K}^{-*} {\rho}^+$ they are given by:
\begin{eqnarray}
H_{00}(D^0 \rightarrow {K}^{-*} \rho^+) & = & - i {{G}_{F} \over 
\sqrt{2} }V_{cs}V_{ud}^{*}m_\rho f_\rho (m_D + m_{{K}^{-*}}) a_1\left\{ 
\alpha A_1^{DK^*}(m_\rho^2) - \right. \nonumber \\
&& \left. \beta A_2^{DK^*}(m^2_\rho) \right\}   \label{eq:a00} \\
H_{\pm\pm}(D^0 \rightarrow K^{*-} \rho^+ ) & = &  i{{G}_{F} \over 
\sqrt{2} }V_{cs}V_{ud}^*m_\rho f_\rho (m_D + m_{K^{*-}}) a_1\left\{ 
A_1^{DK^*}(m_\rho^2) \mp \right. \nonumber \\
&& \left. \gamma V(m^2_\rho)^{DK^*} \right\}.
\label{eq:apm}
\end{eqnarray}
where $\alpha$, $\beta$ and $\gamma$ are function of $r$ and $t$ and 
given by:
\begin{equation}
 \alpha = {1 -{r}^{2} - {t}^{2} \over 2rt}, ~~~~\beta = {k^2 \over 
2rt(1+r)^2} ~\mbox{ and} ~~~ \gamma={k \over (1+r)^2} 
\end{equation}
with $r, t$ and $k$ defined as follow:
\begin{equation} 
 r = {m_{K^*} \over m_D},~~ t = {m_\rho  \over m_D},~~  k^2 = (1 + r^4 + 
t^4 - 2r^2 - 2t^2 - 2r^2t^2).
\end{equation}
For $D\rightarrow K^*\rho$ the values of the parameters $\alpha$, $\beta$ and $\gamma$ are
\begin{equation}
\alpha=1.52,~~~\beta=0.24,~~~\gamma=0.24.
\end{equation}

Equivalently one can work with the partial wave amplitudes which are 
related to the helicity  amplitudes by \cite{ref:digh, ref:nita} ,
\begin{equation}
H_{00} = -{1 \over \sqrt{3} }S +\sqrt{{2 \over 3}}D,~~~~~~~H_{\pm\pm} = 
{1 \over \sqrt{3}}S \pm {1 \over \sqrt{2} } P + {1 \over \sqrt{6}}D.
\label{eq:0pm}
\end{equation}
The partial waves are in general complex and can be expressed in terms 
of their phases as follow
\begin{equation}
S = \mid S\mid\exp{i\delta_S} ,~~P = \mid P\mid \exp{i\delta_P},~~ D = 
\mid D\mid \exp{i\delta_D}.
\label{eq:spd}
\end{equation}
For completeness, we introduce here the transversity basis, $A_{0}$, 
$A_{\parallel}$ and $A_{\perp}$, through
\begin{eqnarray}
A_{0} & = & H_{00}  =  -\sqrt{{1 \over 3}} S +\sqrt{{2 \over 3}}D 
\nonumber \\
A_{\parallel} & = & \sqrt{{1 \over 2}}  (H_{++} + H_{--})  =  \sqrt{{2 
\over 3}}S +\sqrt{{1 \over 3}} D \nonumber \\
A_{\perp} & = & \sqrt{{1 \over 2}}  (H_{++} - H_{--})  =  P.
\end{eqnarray}   
The longitudinal polarization is defined by the ratio of the 
longitudinal decay rate to the total decay rate
\begin{equation}
{P}_{L} = {{\Gamma}_{00} \over \Gamma} = {\mid H_{00}\mid^2 \over \mid 
H_{++}\mid^2 + \mid H_{--}\mid^2 +\mid H_{00}\mid^2}.
\label{eq:pl2}
\end{equation}
Using equations (\ref{eq:a00}),  (\ref{eq:apm}) and (\ref{eq:0pm}) to 
solve for $S$, $P$ and $D$ in term of form factors, we obtain ( we drop a 
common factor of $i {{G}_{F} \over \sqrt{2} }V_{cs}V_{ud}^{*}m_\rho 
f_\rho (m_D + m_{{K}^{-*}}) a_1$ ):
\begin{eqnarray}
S&=&{1 \over \sqrt{3} }\left\{ (2+\alpha)A_1(q^2) - \beta A_2(q^2) 
\right\}, ~~~~P=-\sqrt{2}\gamma V(q^2) ~\mbox{and}  \nonumber \\
D&=&\sqrt{{2 \over 3}} \left\{ (1-\alpha)A_1(q^2) + \beta A_2(q^2) \right\}
\label{eq:spdf} 
\end{eqnarray}
These real amplitudes are assumed to be the magnitudes of the partial 
wave amplitudes. The phases are then fed in by hand. The decay rate given by an incoherent sum,  $  \Gamma \propto  \left(\mid 
H_{++}\mid^2 + \mid H_{--}\mid^2 + \mid H_{00}\mid^2\right)  = 
\left(\mid S\mid^2 + 
\mid P\mid^2 + \mid D\mid^2\right) = \left(\mid A_{0} \mid^2 +\mid 
A_{\parallel} \mid^2 + \mid A_{\perp} \mid^2\right)$, is 
independent of the partial-wave phases. However, the polarization does depend on the phase difference, $\delta_{SD} = \delta_S - \delta_D$, arising from the interference 
between $S$- and $D$-waves,
\begin{equation}
P_L = {1 \over 3} {\mid S\mid ^2 + 2 \mid D\mid^2 - {2\sqrt{2} }\mid 
S\mid\mid D\mid \cos{\delta_{SD}} \over \mid S\mid^2 + \mid P\mid^2 + 
\mid D\mid^2}.
\label{eq:pl}
\end{equation}

The knowledge of the different forms factors is required to proceed 
further with the numerical estimate of the decay rate, $\Gamma$, and the 
longitudinal polarization $P_L$. Since it is not yet possible to obtain the $q^2$ 
dependence of these form factors from experimental data, and a rigorous theoretical
calculation is still lacking, we have relied on several theoretical models for the form 
factors in our analysis. They are: i) Bauer, Stech and Wirbel (BSWI) 
\cite{ref:bsw1}, where an infinite momentum frame is used to calculate 
the form factors at $q^2 = 0$, and a monopole form (pole masses are as 
in \cite{ref:bsw1} ) for $q^2$ dependence is assumed to extrapolate all 
the form factors to the desired value of $q^2$; ii) BSWII \cite{ref:neubert} is a 
modification of BSWI, where while $F_0(q^2)$ and $A_1(q^2)$ are the 
same as in BSWI, a dipole $q^2$ dependence is assumed for $A_2(q^2)$ 
and $V(q^2)$; iii) Altomari and Wolfenstein (AW) model \cite{ref:aw}, 
where the form factors are evaluated in the limit of zero  recoil, and 
a monopole form is used to extrapolate to the desired value of $q^2$; iv) 
Casalbuoni, Deandrea, Di Bartolomeo, Feruglio, Gatto and Nardulli  
(CDDGFN) model \cite{ref:cdd}, where the form factors are evaluated at 
$q^2 = 0$ in an effective Lagrangian satisfying heavy quark spin-flavor 
symmetry in which light vector particles are introduced as gauge 
particles in a broken chiral symmetry.  A monopole form is used for the 
$q^2$ dependence. The experimental inputs for this model are from the 
semileptonic decay $ D \rightarrow  K^* l \nu$, and we have used the 
recent experimental values \cite{ref:pdg6} of the form factors 
$A_1^{DK^*}(0) = 0.55 \pm 0.03$,  $A_2^{DK^*}(0) = 0.40 \pm 0.08$ and  $V^{DK^*}(0) = 1.0 \pm 0.2$, and $f_{D} = 194^{+14}_{-10} \pm 10$ MeV\cite{ref:khadra} 
in calculating the weak coupling constants of the model at $q^2 = 0$ 
\cite{ref:cdd} , which are subsequently used in evaluating the required 
form factors ; v) Isgur, Scora, Grinstein and Wise (ISGW) model 
\cite{ref:isgw}, where a non-relativistic quark model is used to 
calculate the form factors at zero recoil and an exponential $q^2$ 
dependence, based on a potential-model calculation of the meson  wave function, is used to 
extrapolate them to the desired $q^2$; vi) Bajc, Fajfer and Oakes (BFO) 
model \cite{ref:bfo}, where the form factors $A_1(q^2)$ and $A_2(q^2)$ are assumed to be flat, and a monopole behavior is assumed for $V(q^2)$;  and finally (vii), a parametrization that uses experimental values (Exp.F) \cite{ref:pdg6} of the form factors at $q^2 = 0$ and extrapolates them using monopole forms.
\vskip 5mm

\begin{center}
{\bf III. RESULTS }
\end{center}
%\vskip 3mm
\begin{center}
{\bf A. Parameters}
\end{center}
For the numerical calculations we use the following values for the CKM 
matrix elements and meson decay constants :
\begin{eqnarray}
V_{cs} &=& 0.974, ~~~~~V_{ud} = 0.975,~~~~~~~~~~~~~ \nonumber \\
f_{\rho} &=&  0.212~GeV, ~~~~~~f_{K^*} = 0.221~ GeV
\end{eqnarray}

In Table \ref{tab:ff} we present the predicted values of the form 
factors in the different models as well as their experimental values \cite{ref:ex}. 
One observes that while the model predictions for the form factors 
$A_1(q^2)$ and $V(q^2)$ are in the range ($0.6 -1$) and ($0.8 -1.6$), respectively, the model-dependence of $A_2(q^2)$ shows a spread over a larger range: ($0.4 - 3.7$). A striking feature of the BFO model \cite{ref:bfo} is the large value of the form factor $A_{2}$, which is incompatible with its experimental determination.
\begin{center}
{\bf B.  $D^0 \rightarrow K^{*-} \rho^+ $}
\end{center}

We calculate the experimental value of  polarization from the listing 
of Ref.  \cite{ref:pdg}:
\begin{equation}
P_L ={\Gamma(D^0 \rightarrow \rho^+   \bar K^{*-}_{longitudinal}) 
\over \Gamma(D^0 \longrightarrow \rho^+   \bar K^{*-})}= {2.9 \pm 1.2 
\over 6.1 \pm 2.4} = 0.475 \pm 0.271 
\end{equation}

 In Table \ref{tab:mod} we have summarized the results for the decay rates $\Gamma$,  logitudinal polarization $P_L$, and partial-wave ratios ${\mid S\mid \over \mid P\mid }$ and ${\mid S\mid \over \mid D\mid }$ in different models.

We note from Table \ref{tab:mod} that the models CDDGFN, BFO, and the 
scheme that uses experimentally measured form factors, predict a decay 
rate within a standard deviation of the central measured value. All 
other models overestimate the rate by several standard deviations.  As 
for the longitudinal polarization, given the freedom of the unknown 
cos${\delta}_{SD}$, all models are able to fit the data. In particular, 
all models except BFO are able to predict the polarization correctly 
for ${\delta}_{SD}=0$; in the BFO model for ${\delta}_{SD}=0$,  $D^0 
\rightarrow K^{*-} \rho^+ $ becomes totally transversely polarized.  
This circumstance arises from the fact that BFO model predicts a large 
$D$-wave contribution, ${\mid S\mid \over \mid D\mid } \approx \sqrt{2}$. It then becomes evident from Eq.  (\ref{eq:0pm}) that ${H}_{00}$ vanishes. All models except 
BFO also display the partial-wave-amplitude hierarchy: $\mid S\mid > 
\mid P\mid > \mid D\mid$; BFO model on the other hand predicts  $\mid S\mid > \mid D\mid >
\mid P\mid$, which we believe is less likely. The reasoning goes as folows: For decays 
close to threshold, one anticipates the $L^{th}$ partial-wave amplitude to behave like 
${(p/\Lambda)}^{L}$, where $p$ is the center of mass momentum and 
$\Lambda$ a mass scale. For $p\sim 0.4~GeV$ and $\Lambda \sim {m}_{D}$, 
one expects the hierarchy $\mid S\mid > \mid P \mid > \mid D \mid$. 
 %\vspace{5 mm}

\begin{center}
 {\bf C.  $D^+ \rightarrow \bar K^{0*} \rho^+ $}
\end{center}
In contrast to the decay mode  $D^0 \rightarrow \bar K^{-*} \rho^+ $, here the data listing \cite{ref:pdg} is at best confused. First, since the longitudinal and/or transverse branching ratios are not listed, it is not possible to calculate the longitudinal polarization. Second, though  in Refs. {\cite{ref:coff},\cite{ref:jon}} the identification of the transversity amplitudes, (${A}_{T}$, ${A}_{L}$ and ${A}_{l=1}$ in the notation of Ref. \cite{ref:coff}) in terms of the partial-wave amplitudes is correct (see Eqns. (20) - (26) of Ref. \cite{ref:coff}), their identification of the partial-wave amplitudes $S$ and $D$ in terms of the Lorentz structure of the decay amplitude is incorrect. In 
Table II, and more succinctly in Eqns. (32) and (34) of Ref. \cite{ref:coff}, $S$-wave amplitude is identified with the Lorentz structure that goes with the form factor ${A}_{1}$, and $D$-wave 
amplitude with that of ${A}_{2}$. In fact, the correct identification of the $S$- and $D$-wave amplitudes given in Eq. (\ref{eq:spdf}) shows that they are both linear superpositions of ${A}_{1}$ and ${A}_{2}$.

With the caveat that the identification of the partial waves in Refs. \cite{ref:coff,ref:jon} is incorrect (note also that the listing of Ref. \cite{ref:pdg} uses these references only), we take the $S$-, $P$- and $D$-wave branching ratios at their face value and calculate the 'experimental' ratios ${\mid S \mid \over \mid P \mid }$ and ${\mid S\mid \over \mid D \mid }$. 

In Table 2, we have shown the calculated decay rate, the longitudinal polarization and the ratios of the partial-wave amplitudes in different models and compared them with the data. The BFO \cite{ref:bfo} model is the only one that reproduces the total rate correctly. This model also 
generates a large $D$-wave amplitude, with the partial-wave hierarchy $\mid S \mid > \mid D \mid > \mid P \mid$. This feature of the BFO model is due to the exceptionally large value of the form factor $A_2$, which is in contradiction with the experimental detarmination of the form factor as shown in Table 1.
\vspace{5 mm}

\begin{center}
 {\bf D.  $D^0 \rightarrow \bar K^{0*} \rho^0$}
\end{center}
Ref. \cite{ref:pdg} lists the branching ratio, and the transverse 
branching ratio. This enables us to calculate the longitudinal polarization from
\begin{eqnarray}
{P}_{L}=1-{P}_{T} & = & 1 - {B(D^0 \rightarrow { \bar 
K^{0*} \rho^0}_{transverse})\over 
B(D^0 \rightarrow \bar K^{0*} \rho^0 )} \nonumber \\
& = & 0.0 \pm~^{0.4}_{0.0}.
\end{eqnarray}
Ref. \cite{ref:pdg} also lists the $S$- and $D$-wave branching ratios. However, our criticism
of these numbers in the previous subsection applies also to $D^0 \rightarrow \bar K^{0*} \rho^0 $ decay. With this caution, we have taken their \cite{ref:pdg} numbers at face value and calculated  the experimental and theoretical ratios of the partial wave amplitudes and listed them in Table 2.

We note from Table 2 that the rate in the BFO model is too low by three standard deviations; the rates predicted in BSWI and BSWII models are 1.5 standard deviations too high, 
while all other models fit the rate within one standard deviation. As for the longitudinal polarization, all models predict a value consistent with the data.  All models also satisfy the 
${\mid S\mid \over \mid P \mid }$ bound, but only the BFO model fits the ${\mid S\mid \over \mid D \mid }$ ratio. This is 
because the BFO model generates a large $D$-wave amplitude.

A final comment: The inconsistency of the data is evident in the listing \cite{ref:pdg} of the total branching ratio and the individual partial-wave branching ratios. We know that the total branching ratio is an incoherent sum of the individual branching ratios in $S$-, $P$-, and $D$-waves. Yet, in the Particle Data Group listing \cite{ref:pdg}, the sum of $S$- and $D$-wave branching ratios exceeds the total. This by itself should cast doubt on the veracity of data.  
\begin{center}
 {\bf IV.  $S$-wave- and $A_1(q^2)$-dominance}
\end{center}
Since $S$-wave and $D$-wave amplitudes are linear superpositions of the form factors ${A}_{1}$ and ${A}_{2}$, see Eq. (\ref{eq:spdf}),  the concept of $S$-wave-dominance is different from that of $A_1$-dominance. All the models we have discussed, with the exception of BFO model \cite{ref:bfo}, predict that S-wave amplitude is the dominant partial wave amplitude. Further, since Ref. \cite{ref:coff} identifies $S \sim {A}_{1}$ and $D \sim {A}_{2}$, we need to look at what is meant by $S$-wave-dominance and contrast it with $A_1$-dominance.

Consider first the concept of $S$-wave dominance. We see from Eq. (\ref{eq:dec}, \ref{eq:0pm}) that in this approximation, $\Gamma \propto \mid S \mid^2$, and $\mid {H}_{00} \mid= \mid {H}_{++}\mid = \mid {H}_{--} \mid =\mid {S \over \sqrt{3} } \mid$. In practice, most of the models predict the $S$-wave amplitude to be roughly an order of magnitude larger than the $D$-wave amplitude. Consequently, $D$ wave would contribute only 1\% to the rate relative to the $S$-wave.  However, it could influence the longitudinal polarization considerably through its interference with the $S$ wave. Depending on the value of 
${\delta}_{SD}$ the interference term could amount to a $30\%$ correction to $P_L$ (see also Ref. \cite{ref:aren}). However, regardless of the exact size of the $D$-wave amplitude, $S$-wave dominance would lead to $P_L\rightarrow {1 \over 3}$, for ${\delta}_{SD}={ \pi\over 2}$.

Consider now the concept of ${A}_{1}$-dominance. From Eqns. (\ref{eq:a00}) and (\ref{eq:apm}), we see that ${H}_{00} \propto \alpha {A}_{1}$ and ${H}_{++} = {H}_{--} \propto {A}_{1}$. With $\alpha = 1.52$, the longitudinal helicity amplitude is the largest, and the longitudinal polarization becomes  
 \begin{equation}
P_L = {\alpha^2 \over 2 + \alpha^2}=0.54,
\end{equation}
in contrast to a value $1/3 $ (with an error from $S-D$ interference) for S-wave dominance. Further, from Eq. (\ref{eq:spdf}), we note that in ${A}_{1}$-dominance,
\begin{equation}
S \propto  {2 + \alpha \over \sqrt{3} }A_1(q^2),~~~D \propto 
 \sqrt{{2 \over 3}}(1- \alpha ) A_1(q^2),
\end{equation}
which makes the $S$-wave amplitude five times larger than the $D$-wave amplitude - not quite what one would term "$S$-wave dominance."
\begin{center}
 {\bf V. CONCLUSION}
\end{center}
We have carried out an analysis of the process $ D \rightarrow K^* \rho$ in terms of the helicity, and partial-wave amplitudes. We used several models for the form factors, as well as their experimental values, when available, from semileptonic decay. A general feature of our calculation is that all the models, with the exception of the BFO model \cite{ref:bfo},  are consistent with the expected threshold behavior $|S| > |P| > |D|$; 
 $BFO$ model, on the other hand, gives $|S| > |D| > |P|$. Even though in most models the $D$-wave amplitude is almost an order of magnitude smaller than the $S$-wave amplitude, it could effect the polarization prediction significantly through $S-D$ interference.

As we see from Table 2, models BSWI, BSWII, AW and ISGW grossly overestimate the rate for ${D}^{0}\rightarrow {K}^{*-}{\rho}^{+}$, while models CDD, BFO, and the model that uses experimental form factor input, more or less agree with the measured rate. For this decay mode, we trust the measurement of the longitudinal branching ratio as the identification of the transversity amplitudes in Ref. \cite{ref:coff} is correct. Due to the large error in $P_L$, and the uncertainty arising from the $S-D$ interference, all models are consistent with the polarization measurement.

For the mode ${D}^{+}\rightarrow {\bar{K}}^{*0}{\rho}^{+}$, all the models, with the exception of the BFO model \cite{ref:bfo}, grossly overestimate the rate. Before one gets the impression that the BFO model does well, we would like to point out that its prediction for the form factor $A_2$ is in sharp disagrement with the measurements from the semileptonic decays. There are no direct measurements of the longitudinal (or transverse) polarization for this mode. The predicted values of the polarization in every model is almost the same as for the mode ${D}^{0}\rightarrow {K}^{*-}{\rho}^{+}$.

For the mode $D^0\rightarrow {\bar{K}}^{*0}{\rho}^{0}$, BSWI and BSWII models predict a rate within 1.5 standard deviations. The remaining models, with the exception of the BFO model, predict a rate in agreement with data within one standard deviation. BFO model  underestimates the rate by three standard deviations. The transverse rate has been measured \cite{ref:pdg}, from which we have calculated the longitudinal polarization. The measured value of $P_L$ has large errors, but it is consistent with the longitudinal polarization in ${D}^{0}\rightarrow {K}^{*-}{\rho}^{+}$. Given the freedom of the $S-D$ interference, all models  are consistent with the measured polarization. The predicted longitudinal polarization is almost decay-mode independent.

 Final comment: because of the misidentification of the $S$- and $D$-waves with the Lorentz structures in \cite{ref:coff,ref:jon}, we don't trust the partial-wave branching ratios listed in \cite{ref:pdg}. For this reason the listings of ${\mid S\mid \over \mid P \mid }$ and ${\mid S\mid \over \mid D \mid }$ ratios in the last column of Table 2 have to be read with this caveat.

\begin{table}
\centering
\caption{Model and experimental predictions for the form factors : ${{A}_{1,2}}^{DK^* 
(\rho)}(m_{\rho}^2 (m_{K^*}^2)), V^{DK^* (\rho)}(m_{\rho}^2 (m_{K^*}^2))$ and the ratios $x ={A_2(0) \over A_1(0)}, y = {V(0) \over A_1(0)}$ for the process $D \longrightarrow K^* \rho $}
\vspace{5 mm}
\begin{tabular}{|c|c c c c c c c|} \hline
&$BSI$&$BSII$&$AW$&$CDD$&$ISG$&BFO&Exp.F \cite{ref:ex} \\ \hline
$A_1^{DK^*}(m_{\rho}^2)$&0.969&0.969&0.887&0.606&0.909&0.578&0.606 \\
$A_2^{DK^*}(m_{\rho}^2)$&1.264 &1.392&0.707&0.441&0.929&3.747&0.441 \\
$V^{DK^*}(m_{\rho}^2)$&1.414&1.630&1.602&1.153&1.25&0.773&1.153 \\
$x ={A_2(0) \over A_1(0)}$&1.30 &1.30&0.80&0.73&1.02&6.5&0.73 \cite{ref:aitala} \\
$y = {V(0) \over A_1(0)}$&1.39 &1.39&1.73&1.82&1.38&1.16&1.87 \cite{ref:aitala} \\ \hline
\hline
$A_1^{D\rho}(m_{K^*}^2)$&0.898&0.898&0.835&0.732&0.766&0.605&0.637 \\
$A_2^{D\rho}(m_{K^*}^2)$&1.070&1.240&0.846&0.487&0.958&3.574& 0.464\\
$V^{D\rho}(m_{K^*}^2)$&1.529&1.908&1.343&1.326&1.41&0.713&1.248 \\ \hline
\end{tabular}
\label{tab:ff}
\end{table} 
\begin{table}
\caption{Decay rates for $D^{+,0} \longrightarrow \bar {K}^{0*}  {\rho}^{+,0} $.  The values of $\Gamma$ must be multiplied by $10^{11}s^{-1}$. The parameter $z=\cos{\delta_{SD}}$. The experimental values of $P_L$ are listed only if measurements of longitudinal or transverse branching ratios are available \protect \cite{ref:pdg}.} 
\vspace{ 5 mm}
\begin{tabular}{|c|c|c|c|c|c|c|c|c|c|} \hline
& &$BSI$&$BSII$&$AW$&$CDD$&$ISG$&$BFO$&Exp.F\cite{ref:ex}&Expt. \\ \cline{2-10}
$D^0 $&$\Gamma$&4.99&4.96&4.63&2.20&4.56&1.03 &2.20&$1.47\pm .58$ \\ \cline{2-10}
&$P_L$&0.319 - &0.313 - &0.316 -&0.315 -&0.324 -&0.418 - &0.315 -&$0.475 \pm $ \\ 
$\downarrow $& &.084z&.071z &.122z &.127z &.108z&.417z&.127z &0.271  \\ \cline{2-10}
$K^{-*}$&${|S| \over |P|}$&4.3&3.7&3.6&3.5&4.7&2.9&3.5&\\
${\rho}^+ $&${|S| \over |D|}$&10.6&12.3&7.0&6.7&8.3&1.4&6.7&\\ \hline 
$D^+$&$\Gamma$&1.56&1.54&1.50&0.409&1.69&0.268 &0.559&$0.20\pm0.12$ \\ \cline{2-10}
&$P_L$&.326 - &0.325 - &0.319 -&0.318 -&0.333 -&0.416 - &0.321 -&  \\ 
$ \downarrow$&&.086z &.079z &.141z &.128z &.129z&.416z &0.134z &  \\ \cline{2-10}
$  \bar {K}^{0*}$&${|S| \over |P|}$&5.5&5.3&3.6&3.7&6.9&2.8&4.0&$> 2\cite{ref:star}$\\
$\rho^+$&${|S| \over |D|}$&10.6&11.5&6.1&6.7&7.0&1.4&6.5&$1.3 \pm 0.8 \cite{ref:star}$\\\hline 
&$\Gamma $ &$.481$&$.488$&$.426$&$.353$&$.351 $&0.124&0.267&$.354\pm 
.080$ \\ \cline{2-10}
$D^0$&$P_L$&.309 - &.294 - &.314 -&.313 -&.379 -&0.420 - &.307 -&$0.0_{-0}^ {+0.4} $\\ 
$\downarrow$&&.080z &.060z &.097z &.125z &.074z&.419z&.119z &  \\ \cline{2-10}
$\bar {K}^{0*}$&${|S| \over |P|}$&3.4&2.7&3.6&3.3&3.1&3.0&3.0&$>2.8 \cite{ref:star} $\\
${\rho}^0 $&${|S| \over |D|}$&10.7&13.7&8.9&6.8&11.5&1.4&7.1&$1.21 \pm0.23 \cite{ref:star}$\\ \hline
\end{tabular}
\label{tab:mod}
\end{table}

{\bf Acknowledgments}:
This research was partially funded by the Natural Sciences and Engineering Research
Council of Canada through a grant to A. N. K.

\pagebreak


\begin{thebibliography}{99}
\bibitem{ref:ex} The value of experimental form factors ( Exp. F) are caculated using the 
nearest pole approximation: $ F^{DK^*}(m^2)={ F^{DK^*}(0)\over 1-{m^2 \over \Lambda^2}}$ where the values of $F^{DK^*}(0)$ are taken from \cite{ref:pdg6} and the pole masse 
$\Lambda$ are from \cite{ref:bsw1}. In calculating $F^{D\rho^+}(m_{K^*}^2)$ we have used the approximation $F^{D\rho^+}(0) \approx F^{DK^*}(0)$.
\bibitem{ref:star} These values represent numbers extracted from \cite{ref:pdg}. See, however, our criticism of the data.
\end{thebibliography}
\end{document}